\documentclass[12pt]{article}

\usepackage{cite}
\usepackage{epsfig}
\usepackage{amsmath}
\usepackage{amssymb}
\usepackage{authblk}
\usepackage{color}
\usepackage{ulem}

\def\mathswitch#1{\relax\ifmmode#1\else$#1$\fi}
\def\mathswitchr#1{\relax\ifmmode{{#1}}\else${#1}$\fi}
\newcommand{\PW}{\mathswitchr W}
\newcommand{\PZ}{\mathswitchr Z}
\newcommand{\PH}{\mathswitchr H}
\newcommand{\Pe}{\mathswitchr e}
\newcommand{\Pb}{\mathswitchr b}
\newcommand{\Pt}{\mathswitchr t}

\newcommand{\Ps}{\mathswitchr s}
\newcommand{\MW}{\mathswitch {M_\PW}}
\newcommand{\MZ}{\mathswitch {M_\PZ}}
\newcommand{\GZ}{\mathswitch {\Gamma_\PZ}}
\newcommand{\MH}{\mathswitch {M_\PH}}

\newcommand{\me}{\mathswitch {m_\Pe}}
\newcommand{\mb}{\mathswitch {m_\Pb}}
\newcommand{\mt}{\mathswitch {m_\Pt}}

\newcommand{\seff}[1]{\sin^2\theta_{\rm eff}^{#1}}
\newcommand{\al}{\mathswitch {\alpha}}
\newcommand{\as}{\mathswitch {\alpha_\Ps}}
\newcommand{\at}{\mathswitch {\alpha_\Pt}}

\newcommand{\uncer}{{uncertainty}}
\newcommand{\uncers}{{uncertainties}}

\newcommand{\gev}{\,\, \mathrm{GeV}}
\newcommand{\mev}{\,\, \mathrm{MeV}}

\def\order#1{\ensuremath{{\cal O}(#1)}}

\definecolor{orange}{rgb}{0.8,0.5,0}
\definecolor{darkgreen}{rgb}{0,0.5,0}

\begin{document}

\title{\vspace{-2em}Theoretical uncertainties for electroweak and Higgs-boson precision
measurements at FCC-ee}

\setcounter{footnote}{1}
\renewcommand{\thefootnote}{\fnsymbol{footnote}}
\footnotetext[1]{Conveners of Pheno WG2 (``Precision electroweak calculations'')
of the FCC-ee design study}
\renewcommand{\thefootnote}{\arabic{footnote}}
\author[1]{A.~Freitas\textsuperscript{\fnsymbol{footnote}}}
\author[2]{S.~Heinemeyer\textsuperscript{\fnsymbol{footnote}}}
\author[3]{M.~Beneke}
\author[4]{A.~Blondel}
\author[5]{S.~Dittmaier}
\author[6,7]{J.~Gluza}
\author[8]{A.~Hoang}
\author[9]{S.~Jadach}
\author[10]{P.~Janot}
\author[11]{J.~Reuter}
\author[6,12]{T.~Riemann}
\author[13]{C.~Schwinn}
\author[8]{M.~Skrzypek}
\author[14]{S.~Weinzierl}
\affil[1]{\small University of Pittsburgh, Pittsburgh, PA 15260, USA}
\affil[2]{\small Instituto de F\'isica Te\'orica, UAM Cantoblanco, E--28049
  Madrid, Spain;
Campus of International Excellence UAM+CSIC, Cantoblanco, E--28049, Madrid,
Spain; 
Instituto de F\'isica de Cantabria (CSIC-UC), 
E--39005, Santander, Spain}
\affil[3]{\small Physik Department T31, TU M\"unchen, D--85748 Garching, Germany}
\affil[4]{\small University of Geneva, Geneva, Switzerland}
\affil[5]{\small Freiburg University, D--79104 Freiburg, Germany}
\affil[6]{\small Institute of Physics, University of Silesia, Katowice,
  Poland}
\affil[7]{Faculty of Science, University of Hradec Kr\'alov\'e, Czech Republic}
\affil[8]{\small University of Vienna, A--1090 Wien, Austria}
\affil[9]{\small Institute of Nuclear Physics, PAN, 31-342 Krakow, Poland}
\affil[10]{\small CERN, PH Department, Geneva, Switzerland}
\affil[11]{\small DESY Theory Group, D--22607 Hamburg, Germany}
\affil[12]{\small DESY Theory Group Zeuthen, D--15738 Zeuthen, Germany}
\affil[13]{\small RWTH Aachen, D--52056 Aachen, Germany}
\affil[14]{\small PRISMA Cluster of Excellence, University of Mainz, D--55099
Mainz, Germany}

\date{}

\maketitle


\mbox{}\vspace{-3.5em}
\begin{abstract}
\noindent
Due to the high anticipated experimental precision at the Future
Circular Collider FCC-ee (or
other proposed $e^+e^-$ colliders, such as ILC, CLIC, or CEPC)  for
electroweak and Higgs-boson precision measurements, theoretical
uncertainties may have, if unattended, an important impact on the
interpretation of these measurements within the Standard Model (SM), and 
thus on constraints on new physics. 
Current theory uncertainties, which would dominate the total
uncertainty, need to be strongly reduced through future 
advances in
the calculation of multi-loop radiative corrections together with improved
experimental and theoretical control of the precision of
SM input parameters.
This document aims to provide an estimate of 
the required improvement in calculational accuracy in view of the anticipated high
precision at the FCC-ee. For the most relevant electroweak and
Higgs-boson precision observables we evaluate the corresponding
quantitative impact.\\[.6em] 
{\tt IFT-UAM/CSIC-18-021, TUM-HEP-1185/19, KW 19-001, TTK-19-20, UWThPh 2019-16,
IFJPAN-IV-2019-8, FR-PHENO-2019-010, DESY 19-105, arXiv:1906.05379}

\end{abstract}


\newpage
\section{Introduction}

With the discovery of the Higgs boson, all elements of the Standard Model have
been experimentally confirmed and tested in great depth. On the other hand,
observational evidence for neutrino masses, dark matter and the
matter-antimatter asymmetry require physics beyond the Standard Model. One
promising way to probe such new physics is through precision measurements of the
properties of the electroweak gauge bosons and the Higgs boson. This is the
avenue pursued by several proposals for a future $e^+e^-$ collider. In
particular, the FCC-ee concept is designed to run at the $Z$ pole, the $WW$
threshold, as a Higgs factory, and at the $t\bar{t}$ threshold, and thus it is
can improve indirect probes for
new physics from all of these by several orders of magnitude compared to existing bounds
\cite{fccee,cdr}.

The anticipated experimental accuracy of an observable has to be matched
with a theory prediction of at least the same level of accuracy to make maximum
use of the experimental data. 
Both types of uncertainties (theoretical and experimental) must be taken into
account when deriving constraints on new physics from the data.

Several sources of theory uncertainties have to be distinguished. 
The {\em intrinsic} uncertainties are due to missing higher-orders in
the perturbative expansion of the SM (or BSM) prediction for an observable. 
The {\em parametric} uncertainties
are due to the imperfect experimental knowledge of the SM input
parameters as well as theory uncertainties induced in their extraction from 
data\footnote{The distinction between observables and input
parameters is somewhat arbitrary in global electroweak fits. However, if an
observable is affected by a large uncertainty from an input parameter,
the precision of the global fit will typically be impacted to a similar degree.
For this reason we will refrain from discussing global fits separately in this
document.}. The extraction of a quantity from a cross section or an
asymmetry requires the theory prediction of this cross section or
asymmetry to at least the same order of precision\footnote{It is worth noting
that in some cases the
impact of theory uncertainties can be reduced by analyzing certain ratios or
difference of measured quantities.}.
Finally, it is worth mentioning that the achievable precision for the
prediction of a given observable is not only limited by the theory
uncertainty to this observable. In many cases, the determination 
of the ``observable'' itself from experimental data requires theory input for
the subtraction of background and/or the evaluation of the impact of
experimental acceptances. Thus the typical electroweak precision
observables and Higgs precision observables are technically
``pseudo-observables''.

In this paper the current status and future implications of all three
types of theory uncertainties will be summarized. 
While they have been obtained for the FCC-ee, they are valid for
all future high precision $e^+e^-$ colliders running on the $Z$ pole or
above the $HZ$ threshold, such as 
ILC, CLIC, or CEPC. We will use anticipated FCC-ee precisions
to illustrate the impact of theory uncertainties.
This compilation may serve as a
reference for other FCC-ee (or similar) studies.


\section{Methods for estimation of theory errors}

Some of the commonly used techniques for the determination of theory errors are
(for a review see Ref.~\cite{Freitas:2016sty})
\begin{itemize}

\item
Determine relevant prefactors pertaining to a class of higher-order
corrections, such as couplings, group factors, particle multiplicities,
mass ratios, {\it etc.} It is assumed that the remainder of the loop
amplitude is \order{1}.  

\item
If several orders of radiative corrections to some quantity have already
been computed, one can extrapolate to higher orders based on the
observation that the perturbation series approximately follows a
geometric series.%
\footnote{It should be kept in mind that the series coefficients receive
combinatorial enhancement factors at
  higher orders. The method described here is a heuristic ansatz that
  empirically works for extrapolations to two- or three-loop order.}

\item
When using $\overline{\text{MS}}$ renormalization, one can use the scale
dependence of a given fixed-order result to estimate the size of the missing
higher orders. This method is commonly used in the context of QCD corrections,
but is less useful for electroweak calculations.

\item
Compare the results in different renormalization schemes, such as the on-shell
and $\overline{\text{MS}}$ schemes, where the differences are of the next
order in the perturbative expansion.

\end{itemize}
There is no formal argument concerning the validity of any of these methods
and their usefulness is mostly supported based on experience. The estimates in
the following subsections are mostly derived using the prefactor and geometric
series methods. The reader should take the numbers presented below with a grain of salt and be
aware that they are educated guesses rather than robust quantitative
predictions.


\section{SM parameter determination}

The prediction of precision (pseudo-)observables depends on
the input quantities $G_{\rm F}$,
$\alpha(\MZ) \equiv \alpha/(1-\Delta\alpha)$, $\as(\MZ)$, $\MZ$, $\MH$, 
$\mt$ and $\mb$ (and $m_{\rm c}$). 
These need to be determined with sufficient precision from independent
measurements. $G_{\rm F}$ is currently known with a very small
uncertainty of 0.5 ppm, such that it will be a subdominant \uncer\ source
for future FCC-ee precision tests.
All of the remaining parameters can, in principle, be measured 
at FCC-ee
with improved precision compared to current world averages. However, each of
these measurements is also subject to theory uncertainties itself.
Here we list the future expectations for the most relevant SM input
parameters, as derived for the FCC-ee. For other $e^+e^-$ collider
experiments different uncertainties are expected, and the relevant numbers
given below could be rescaled.

\smallskip
\underline{$\MH$:} 
The Higgs mass can be measured from $e^+e^- \to HZ$ at $\sqrt{s}=240\gev$ with a
precision of the order of 10~MeV \cite{Azzi:2012yn,Moortgat-Picka:2015yla}. At this level of
precision, theory uncertainties (mainly due to final-state radiation effects)
are subdominant. 

\smallskip
\underline{$\MZ$:}
The $Z$ mass can be determined from the line-shape of the process $e^+e^- \to
f\bar{f}$ at several center-of-mass energies above and below $\sqrt{s}=\MZ$. The
experimental precision at FCC-ee is estimated to be below 0.1~MeV. On the theory
side, this measurement is primarily affected by QED initial-state radiation and
interference between initial and final-state. Smaller modification arise
from non-resonant photon 
exchange and electroweak box diagrams. In Ref.~\cite{Jadach:1999pp} it was estimated that the uncertainty
for the current best calculations of these effects amounts to less than 0.1~MeV.

\smallskip
\underline{$\as(\MZ)$:}
Given the overconstrained system of electroweak precision pseudo-observables,
the value of $\as(\MZ)$ can be extracted from a global fit to these quantities.
The experimental uncertainty from such a fit to FCC-ee data is projected
to be about $10^{-4}$ \cite{alphas}.
One advantage of this approach is that the strong coupling is obtained directly
at the scale $\mu=\MZ$, where non-perturbative uncertainties are suppressed by
powers of $\MZ^2$, in contrast to $\as$ determinations from hadronic $\tau$
decays or jet event shapes. However,
perturbative uncertainties will be important. To evaluate their impact, one can
look at $R_\ell$, which is one of the most sensitive pseudo-observables for
$\as$. With $\delta_{\rm th}R_\ell = 1.5\times 10^{-3}$, see 
Tab.~\ref{tab:futerr} below, one obtains a theory uncertainty of
0.00015 for $\as(\MZ)$ in addition to the experimental uncertainty.

\smallskip
\underline{$\mt$} (defined as the $\overline{\text{MS}}$ mass, $\mt(\mt)$):
The top-quark mass can be determined from a threshold scan near the pair
production threshold at $\sqrt{s} \approx 350\gev$, using a threshold
mass definition (1S or PS). The projected experimental
uncertainty for this measurment is 
$\delta \mt \approx 17\mev$~\cite{cdr}. 
On the theory side, there are several \uncer\ sources \cite{Vos:2016til}: 
\emph{(i)} The perturbative uncertainty
for the calculation of the threshold shape. The currently most precise
result includes NNNLO QCD~\cite{beneke} and NNLO EW and non-resonant
corrections (counting orders in non-relativistic, resummed perturbation theory
for the top threshold)~\cite{Beneke:2017rdn}, resulting in an
error of $\mt$ of somewhat less than 50~MeV.
It is expected that this can be
further improved by matching the fixed-order calculation with resummed
calculations based on effective field 
theories, currently available at the NNLL level \cite{Hoang:2013uda},
but it may be difficult to reduce the error by a factor of two or
more. \emph{(ii)} The threshold shape calculations are performed by using a
threshold mass definition (1S or PS), which must be translated into the
$\overline{\text{MS}}$ scheme. This translation is currently known to the
four-loop level \cite{marquard}, resulting in an uncertainty of about 10~MeV.
\emph{(iii)} The threshold shape prediction and mass scheme translation depend
on the value of $\as$. Assuming an uncertainty $\delta\as \sim 0.001$ leads to a
top-quark mass error of  $\delta\mt \sim 15\mev$ for the measurement of a top
threshold mass definition~\cite{Beneke:2017rdn}, but to an uncertainty in the
conversion to the $\overline{\text{MS}}$ mass of roughly $70\mev$
\cite{Hoang:2000yr}. With a future improved  measurement of $\as$ at FCC-ee (see
above), the latter error contribution  can also be reduced to the level of
$15\mev$. Combining these
three error sources, a theory uncertainty of less than 50~MeV for $\mt$ appears
feasible.

Besides the issues discussed above, the achievement of this goal will require
a very accurate determination of the efficiencies of experimental acceptances
and selection cuts. This task is facilitated by the inclusion of higher-order
corrections and resummation results in a Monte-Carlo event generator. Work in
this direction is underway, with recent evaluations the NNLL resummed
threshold-shape in the presence of phase-space cuts
\cite{Hoang:2013uda}, NLO QCD corrections for off-shell 
$t\bar{t}$ production \cite{Chokoufe:2016bci}, and matching between these
contributions \cite{Reuter:2016ohp}, complementing the semi-analytic
approach~\cite{Beneke:2010mp,Jantzen:2013gpa} implemented in the most 
recent NNNLO QCD + NNLO EW/non-resonant calculation \cite{Beneke:2017rdn}.

\smallskip
\underline{$\mb$ and $m_{\rm c}$} (defined as $\overline{\text{MS}}$ masses,
$\mb(\mb)$ and $m_{\rm c}(m_{\rm c})$):
We use estimated future values of $\delta \mb(\mb) \sim 13\mev$ and 
$\delta m_{\rm c}(m_{\rm c}) \sim 7 \mev$ \cite{lmp}. These
are based on projected improvements in lattice calculations, for which we take
the moderately conservative LS scenario in Ref.~\cite{lmp}. Note that some
analysis based on QCD sum rules already claim an uncertainty of $\delta\mb \sim
10\mev$~\cite{mbsum}, but these error estimates are not confirmed in 
other analyses of the same quantities~\cite{Chetyrkin:2009fv, mbc}.
It would be very welcome to have the two independent results from
lattice and QCD sum rules for cross-checking and for putting uncertainty
estimates to a more solid basis.

\smallskip
\underline{$\Delta\al$:}
The potentially most difficult parameter to measure is the shift in the
electromagnetic fine structure constant, $\Delta\alpha \equiv
1-\alpha(0)/\alpha(\MZ)$. It is traditionally determined from data on $e^+e^-
\to \text{hadrons}$ and tau decays to hadrons~\cite{dalpha}, with a current
uncertainty of $\delta(\Delta\alpha) = \order{10^{-4}}$. More accurate data for
these inputs is expected to become available from BES~III~\cite{bes}, VEPP-2000
\cite{vepp} and
Belle~II~\cite{belle}. It was estimated that an uncertainty of 
$4 \times 10^{-5}$ to $5 \times 10^{-5}$ could be reached, depending on
improvements in QCD theory input~\cite{fredl}.
At FCC-ee, independent measurements of 
$e^+e^- \to \text{hadrons}$ could be obtained through the radiative
return method~\cite{isr}. The ultimate precision on $\Delta\alpha$ that could be
reached by this method is not yet clear, but in view of the
importance of the quantity $\Delta\alpha$ for EW processes it would be highly
desirable to have independent results on $\Delta\alpha$ for a firm uncertainty
assessment.

In addition, the expected high luminosity may, for the first time,
enable the possibility to measure $\alpha(\MZ)$ directly from $e^+e^- \to
f\bar{f}$ at $\sqrt{s_\pm} = \MZ \pm 3\gev$ \cite{janot}. 
In Ref.~\cite{janot} it was evaluated
that from an analysis of the FB asymmetry of $e^+e^- \to \mu^+\mu^-$ one can
determine $\alpha(\MZ)$ with an experimental \uncer\ of $3 \times 10^{-5}$.
However, this also requires theory input for the subtraction of other
electroweak corrections from the pure $s$-channel photon exchange contribution,
which is the part that directly depends on $\alpha(\MZ)$. In particular, one
needs to subtract contributions from $s$-channel $Z$ exchange, box diagrams, and
corrections to the $\gamma ff$ vertex. All of these contributions are currently
at the one-loop level (see for example Ref.~\cite{zfitter}) and have an impact
of \order{10^{-3}} on the extracted value of $\alpha(\MZ)$. 
This value is relatively small due to partial cancellations between $A_{\rm FB}$
at $s_+$ and $s_-$, but still large compared to the experimental target.
With existing loop
calculation methods, it is possible to compute complete fermionic two-loop
corrections, as well as \order{\alpha\as^2} corrections. After inclusion of
these contribution, the remaining theory uncertainty is estimated to amount to
\order{10^{-4}}. If additionally the full \order{\alpha^2}, \order{\al^2\as}
and double-fermionic \order{\alpha^3} corrections become
available, this \uncer\ may be reduced to the level of a few times $10^{-5}$.
These calculations (which include two- and three-loop box diagrams) 
would require new developments for loop integration
techniques, but may be achievable in the future.

There are also important QED effects in $A_{\rm FB}(s_+)-A_{\rm FB}(s_-)$ due to
initial-final state interference, examined in the recent study of
Ref.~\cite{Jadach:2018jjo}. They are numerically much bigger than non-QED EW
corrections --
amounting to about 0.5\%, in spite of the partial cancellation in the difference.
It was shown in Ref.~\cite{Jadach:2018jjo}, that thanks to an
advanced technique for soft photon resummation, 
it can be controlled theoretically with a precision of \order{0.01\%}.
Another factor $\sim 10$ improvement is needed in order to match FCC-ee 
experimental precision of the direct measurement of $\alpha(\MZ)$.

For illustration, two scenarios for $\Delta\alpha$ will be considered below, 
one with total anticipated uncertainty of 
$5 \times 10^{-5}$ (assuming a combination of the experimental and
possible future theory uncertainties of similar magnitude) and an optimistic
one with total 
uncertainty of $3 \times 10^{-5}$ (corresponding to subdominant theory
uncertainties). For the optimistic scenario, we also consider a reduced
uncertainty of $\as$, which may be achievable by combining several observables
\cite{dEnterria:2018cye}.

\vspace{\medskipamount}
Taking into account the experimental and theoretical \uncers\ discussed above,
one 
arrives at the following estimates for the achievable precision 
(from direct determination) for the most important SM parameters at the
  FCC-ee: 
\begin{align}
\delta\mt &= 50 \mev, \quad
\delta\mb = 13 \mev, \quad
\delta\MZ = 0.1 \mev, \quad
\delta\as = 0.0002\; (0.0001), \quad \nonumber \\
\delta(\Delta\alpha) &= 5\times 10^{-5} \; (3\times 10^{-5}).
\label{eq:paraerr} 
\end{align}
For $\mt$ and $\as$, another factor two improvement could be envisioned with a
more ambitious theory advancement.


\section{Electroweak precision observables}

\subsection{EWPO definitions}

The most important electroweak precision observables (EWPO) are related
to properties 
of the $Z$ and $W$ bosons. $Z$-boson properties are determined from measurements
of $e^+e^- \to f\bar{f}$ on the $Z$-pole. To isolate the physics of the
$Z$-boson, the typical set of pseudo-observables is defined in the terms of the
de-convoluted cross-section $\sigma_f(s)$, where the effect of initial- and
final-state photon radiation and from $s$-channel photon and double-boson (box)
exchange  has been removed. The impact of these corrections will be discussed
below. The customary set of pseudo-observables are
\begin{align}
\sigma^0_{\rm had} &= \sum_q \sigma_q(\MZ^2), \\
\GZ &= \sum_f \Gamma[Z\to f\bar{f}], \qquad \text{(from a
fit to $\sigma_f(s)$ at various values of $s$)} \\
R_\ell &= \bigl [{\textstyle\sum_q \sigma_q(\MZ^2)}\bigr ]/\sigma_\ell(\MZ^2), \qquad
(\ell = e,\mu,\tau) \\[1ex]
R_q &= \sigma_q(\MZ^2)/\bigl [{\textstyle\sum_q \sigma_q(\MZ^2)}\bigr ], \qquad
(q = b,c) \displaybreak[0] \\[1ex]
A^f_{\rm FB} &= \frac{\sigma_f(\theta<\frac{\pi}{2})-
 \sigma_f(\theta>\frac{\pi}{2})}{\sigma_f(\theta<\frac{\pi}{2})+
 \sigma_f(\theta>\frac{\pi}{2})} \equiv \tfrac{3}{4}{\cal A}_e{\cal A}_f, 
 \label{eq:afb} \displaybreak[0]\\[1ex]
A^f_{\rm LR} &= \frac{\sigma_f(P_e<0)-
 \sigma_f(P_e>0)}{\sigma_f(P_e<0)+
 \sigma_f(P_e>0)} \equiv {\cal A}_e|P_e|.
\end{align}
Here $\theta$ is the scattering angle and $P_e$ is the polarization of the
incoming electron beam.%
\footnote{Formulas for electron {\em and} positron polarization can be found, e.g., in
Ref.~\cite{MoortgatPick:2005cw}.}  
~The asymmetry parameters are commonly written as
\begin{align}
{\cal A}_f &= \frac{1-4|Q_f|\seff{f}}{1-4|Q_f|\seff{f}+8 (Q_f\seff{f})^2}.
\end{align}
Here $Q_f$ denotes the charge of the fermion, and $\seff{f}$ is the 
effective weak (fermionic) mixing angle.
Another important precision observable is the $W$-boson mass. It is currently
measured most precisely from the lepton $p_\perp$ distribution in $pp \to
\ell\nu$ at hadron colliders, and it can be calculated within the SM from the
Fermi constant, $G_{\rm F}$, of muon decay.

\vspace{\medskipamount}

The expected precision for the experimental determination of some of these
quantities at FCC-ee is given in the second column of
Tab.~\ref{tab:futerr}~\cite{fccee,cdr,d'Enterria:2016yqx}. 
The $Z$-boson quantities can be determined from a run at $\sqrt{s}=\MZ$ with
several ab$^{-1}$, and smaller statistics runs at center-of-mass energies above
and below the $Z$ peak for the purpose of $\MZ$ and $\GZ$ measurements. The $W$
mass can be determined from a run at several values of $\sqrt{s}$ near the
threshold $2\MW$
with a combined luminosity of ${\cal O}$(ab$^{-1}$).
Note that
the number for $\MW$ in the table includes an estimate of the theory error as described in
section~\ref{sec:ewpo-para}, since the measurement of $\MW$ requires a full SM
prediction (not only QED) for the $WW$ cross-section near threshold as input.

\begin{table}[tb]
\centering
\renewcommand{\arraystretch}{1.1}
\begin{tabular}[t]{l|cclc}
\hline
Quantity & FCC-ee & \multicolumn{2}{l}{Current intrinsic error} & Projected intrinsic error \\
\hline
$\MW$ [MeV] & 0.5--1{$^{\,\ddagger}$} & 4 & ($\al^3, \al^2\as$) & 1 \\
$\seff{\ell}$ [$10^{-5}$] & 0.6 & 4.5 & ($\al^3, \al^2\as)$ & 1.5 \\
$\GZ$ [MeV] & 0.1 & 0.4 & ($\al^3, \al^2\as, \al\as^2$) & 0.15 \\
$R_b$ [$10^{-5}$] & 6 & 11 & ($\al^3, \al^2 \as)$ & 5 \\
$R_l$ [$10^{-3}$] & 1 & 6 & ($\al^3, \al^2 \as$) & 1.5 \\
\hline
\multicolumn{5}{l}{{$^\ddagger$The pure experimental precision on
    $\MW$ is $\sim 0.5 \mev$~\cite{fccee,cdr}, see Sec.~\ref{sec:ewpo-para}
      for more details.}}
\end{tabular}
\caption{Estimated precision for the direct determination of several
important electroweak precision observables at FCC-ee
\cite{fccee,cdr,d'Enterria:2016yqx} (column two, including systematic
and observable-specific) uncertainties; as well as current intrinsic
theory errors for the prediction of these quantities within
the SM (column
three). The main 
sources of theory errors are
also indicated.
Column four shows the estimated projected intrinsic
theory errors when leading 3-loop corrections become available.
See text for more details.}
\label{tab:futerr}
\end{table}


\subsection{Theory uncertainties for EWPO}

\subsubsection{Intrinsic uncertainties}

The quantities listed in Tab.~\ref{tab:futerr} can be predicted within
the SM by using $G_{\rm F}$, 
$\alpha(\MZ)$, $\as(\MZ)$, $\MZ$, $\MH$ and $\mt$ as inputs.  
The radiative corrections in these predictions
are currently known including complete two-loop corrections
In addition, approximate three- and
four-loop corrections of \order{\at^3}, \order{\at^2\as}, 
\order{\at\as^2} and \order{\at\as^3} are available, where $\at =
y_\Pt^2/(4\pi)$ and $y_\Pt$ is the top Yukawa coupling. For a review, see
Ref.~\cite{Freitas:2016sty}. The theory uncertainties from missing
higher-order corrections are given in the third column of   
Tab.~\ref{tab:futerr}~\cite{therr1}. 
Also indicated are the main sources for the respective uncertainties.

\smallskip
As evident from the table, to match the anticipated FCC-ee precision,
substantial improvements in the SM theory prediction are necessary. In
Ref.~\cite{futth,Blondel:2018mad}, it was estimated how the intrinsic uncertainty will
likely be reduced if the following calculations become available:
complete \order{\alpha\as^2} corrections, fermionic \order{\alpha^2\as}
corrections, double-fermionic \order{\alpha^3} corrections, and 
leading four-loop corrections enhanced by the top Yukawa coupling.

With the inclusion of these corrections, the estimated future intrinsic
uncertainties will become comparable to the anticipated experimental
FCC-ee precision, as shown in the fourth column of
Tab.~\ref{tab:futerr}. However, for some quantities the calculation of
additional terms will be necessary to match the anticipated experimental
accuracies, in particular for $\seff{\ell}$. These extra terms include
subdominant three-loop and leading four-loop contributions\footnote{This future
scenario is called TH2 in Ref.~\cite{Blondel:2018mad}.}.

To carry out these calculations, qualitatively new developments of existing loop
integration techniques will be required, but no conceptual paradigm shift.
An extensive overview of recent progress in loop calculation techniques and prospects for
future developments can be found in Ref.~\cite{Blondel:2018mad,Blondel:2019vdq}. For the
calculation of full SM corrections, numerical integration techniques are often
advantageous due to the large number of independent mass and momentum scales
involved. Some techniques, such as sector decomposition and Mellin-Barnes
representations can in principle be applied to problems with arbitary number of
legs and loops. Several public computer codes based on these methods are
available. Nevertheless, substantial improvements will be needed to ensure
reliable numerical precision of at least 2--3 digits for future 3-loop and
4-loop applications. In some cases where higher precision is needed, more
specialized semi-numerical methods can be applied. On the other hand, for
QED and QCD corrections on external legs (see below), 
analytical techniques are typically more efficient if the radition is treated
inclusively, or Monte-Carlo techniques if experimental cuts are applied to the
phase-space. See
Ref.~\cite{Blondel:2018mad,Blondel:2019vdq} for more details.

\vspace{\medskipamount}

On the other hand, it should be kept in mind that the determination of
the pseudo-observables in 
Tab.~\ref{tab:futerr} from experimental data also requires theory input for the
removal of initial-state and final-state photon radiation and $s$-channel
photon exchange and box contributions. 

For the total cross-section for $e^+e^- \to f\bar{f}$, the contributions
from QED radiation are currently known with complete \order{\al} and 
\order{\al^2} corrections and log-enhanced \order{\al^3 L^3} corrections.
For the asymmetries, \order{\al} and \order{\al^2 L^2} corrections are
available (with $L \equiv \log s/\me^2$). More details can be found, for
example, in Ref.~\cite{pcp}, see also the discussion in Ref.~\cite{bluemlein}. 

The theory uncertainty from missing higher QED orders is estimated to amount to a
few times 0.01\% \cite{pcp,lep1} for the $Z$-peak cross-section and total width
measurements. In order not to be dominated by this uncertainty, 
it will 
need to be reduced by about a factor 10 for the FCC-ee. This will require the
calculation of non-leading log contributions to \order{\al^3}
corrections, \order{\al^3 L^2} and \order{\al^4 L^4} contributions,
as well as an improved treatment of fermion pair production from
off-shell photons.

Generally, the concept of pseudo-observables to describe the $Z$ line shape and
asymmetries should be validated (and modified if
necessary) at the aimed target precision. On the one hand, a full calculation of
the resonance process at ${\cal O}(\alpha^2)$, including all off-shell
contributions and weak corrections, seems necessary. Technically, this is a
challenging two-loop calculation with massive particles; conceptually, a method
to treat resonance processes (including off-shell tails) is required at the
two-loop level. On the other hand, the field-theoretical definition of the
pseudo-observable should be consistently based on the complex $Z$ pole, and the
parametrization of the relevant cross sections via pseudo-observables should be
carefully compared to the best achievable prediction. Though very challenging,
we are confident that both steps can be made in theory.

The role of theory uncertainties in the determination of $\MW$ from a threshold
scan will be discussed in the next subsection.


\subsubsection{Parametric uncertainties}
\label{sec:ewpo-para}

As discussed above, the prediction of electroweak precision pseudo-observables
depends on the input quantities $G_{\rm F}$,
$\alpha(\MZ) \equiv \alpha/(1-\Delta\alpha)$, $\as(\MZ)$, $\MZ$, $\MH$ and
$\mt$. Their anticipated precisions, including future FCC-ee measurements are
summarized in Eq.~\ref{eq:paraerr}. In the following we discuss the impact of
each of these parameters on the EWPO.

\smallskip
\underline{$\MH$:} 
The Higgs mass can be measured with a precision of the order of 10~MeV.
The dependence of other electroweak
precision pseudo-observables on $\MH$ is proportional to $\alpha \log \MH/\MW$
and thus relatively mild, so that an accuracy below 0.05~GeV will lead to a
negligible error contribution for electroweak precision tests.

\smallskip
\underline{$\MZ$:}
The $Z$ mass is expected to be determined to better than 0.1~MeV.
This very small error will not be a limiting factor for electroweak precision fits.

\smallskip
\underline{$\as(\MZ)$:}
With a theory uncertainty of 0.00015 for $\as(\MZ)$ in addition to the
experimental uncertainty, one arrives at the parametric uncertainties as given
in Tab.~\ref{tab:paraerr}.

\smallskip
\underline{$\mt$:}
With a top quark mass uncertainty of $50 \mev$ one arrives at the parametric
uncertainties as given in Tab.~\ref{tab:paraerr}.

\smallskip
\underline{$\mb$:}
The impact of the bottom-quark mass cannot be entirely neglected,
but its uncertainty contribution is negligibly small. 

\smallskip
\underline{$\Delta\al$:}
For illustration, we use two scenarios for $\Delta\alpha$ in our evaluations, 
one with total anticipated uncertainty of 
$5 \times 10^{-5}$ (assuming a combination of the experimental and
possible future theory uncertainties of similar magnitude) and an optimistic
one with total 
uncertainty of $3 \times 10^{-5}$ (corresponding to subdominant theory
uncertainties). The corresponding results can be found in
Tab.~\ref{tab:paraerr}.

\smallskip
\underline{$\MW$:}
Here we also discuss the $W$-boson mass, which will (most likely)
not serve as an input parameter. Similarly to the top quark mass, $\MW$
can be determined from a threshold scan near the $W$-pair 
threshold, $\sqrt{s} \approx 161\gev$. It is forseen that the experimental
uncertainty at FCC-ee for this measurement is about 0.5~MeV \cite{fccee,cdr}. 
At the
point of highest sensitivity, an uncertainty of the cross-section
measurement of $0.1\%$ translates to an uncertainty of $\sim 1.5$~MeV on
$M_W$~\cite{Azzi:2017iih}. Therefore a theoretical prediction for the
process $e^+e^-\to 4f$ with an accuracy of $\Delta\sigma\sim 0.01\%$ is
desirable, including effects of off-shell W bosons, which become
important near threshold.

The currently
best calculations are based on complete one-loop results for $e^+e^- \to 4f$
\cite{Denner:2005fg} and partial higher-order effects 
for the total cross section from an effective field
theory framework \cite{Beneke:2007zg,Actis:2008rb}. The resulting theory
\uncer\ on $\MW$ 
is estimated to be about 3~MeV \cite{Actis:2008rb}. Building on the effective
field theory framework of Ref.~\cite{Beneke:2007zg}, this result could be
improved if complete 2-loop corrections to $e^+e^- \to W^+W^-$ and to $W \to
f\bar{f}'$ become available (which are required for the determination of
matching coefficients in the effective theory). While such a calculation
poses a challenge, it may be feasible with a serious effort over some
number of years. In addition, a more accurate description of
initial-state radiation will be important, which includes universal
contributions from soft and collinear photon radiation (see Ref.~\cite{pcp} for a
review), as well as hard photon radiation. For the latter, a proper matching and
merging procedure needs to be employed to avoid double counting
\cite{Denner:2000bj,Beneke:2007zg}. 
Estimates based on scaling arguments~\cite{Skrzypek:FCCee} or explicit
calculation of a subset of leading, Coulomb enhanced, three-loop
effects~\cite{Schwinn:FCCee} suggest that corrections beyond NNLO are of
the order of $0.02\%$. The full computation of the Coulomb-enhanced
three-loop corrections is expected to be feasible after completion of
the NNLO EFT calculation. The remaining, non Coulomb-enhanced
\order{\alpha^3} corrections are therefore expected to be below
the FCC-ee target accuracy. The magnitude of single- or non-resonant
contributions have been estimated as $0.016$-$0.03\%$
\cite{Skrzypek:FCCee,Schwinn:FCCee}. These corrections would be included
in a full NNLO computation of $e^+e^-\to 4f$, but it may also be
possible to compute them within the EFT. These estimates suggest a
theory induced systematic error $ \Delta \MW = (0.15-0.60)$~MeV, where the
lower value results from assuming the non-resonant corrections are under
control.

Current cross-section predictions for $e^+e^-\to WW\to4f$ involving 
final-state quarks do not yet fully control all
effects of ${\cal O}(\alpha_s^2)$, which is, however, needed to meet the
0.01\%--0.04\% precision tag for the total cross section. 
For QCD corrections of ``factorizable''
origin, which correspond to corrections to hadronic on-shell or off-shell $W/Z$
decays, these corrections are straightforward to supplement, at least for
integrated quantities. ``Non-factorizable'' QCD interconnection corrections,
which involve two-parton exchange between two hadronically decaying $W$ or $Z$
bosons, are more complicated. Their leading resonant contribution should be
calculable, so that the remaining uncertainty on those effects will be ${\cal
O}(\as^2/\pi^2\times\Gamma_\PW/\MW)<0.01\%$. It is even expected
\cite{Fadin:1993kt} that the resonant part of the non-factorizable QCD
interconnection effects vanishes to all orders in $\as$ for inclusive
quantities.\label{wwqcd}

\vspace{\medskipamount}
The impact of the input uncertainties in eq.~\eqref{eq:paraerr} on the
theoretical prediction of the 
pseudo-observables in Tab.~\ref{tab:futerr} is given in Tab.~\ref{tab:paraerr}.
Here the numbers in in brackets refer to the optimistic scenario for the
theory error of $\Delta\alpha$. As evident from the table, the limiting factors
among the input parameters are $\as$ and $\Delta\alpha$. The present estimation
of the parametric errors
can in some cases exceed the anticipated experimental FCC-ee precision, but not by more
than a factor of about 2.

\begin{table}[tb]
\centering
\renewcommand{\arraystretch}{1.1}
\begin{tabular}[t]{l|ccc}
\hline
Quantity & FCC-ee & future parametric unc.\ & Main source \\
\hline
$\MW$ [MeV] & $0.5 - 1$ & 1 (0.6) & $\delta(\Delta\alpha)$  \\
$\seff{\ell}$ [$10^{-5}$] & 0.6 & 2 (1) & $\delta(\Delta\alpha)$ \\
$\GZ$ [MeV] & 0.1 & 0.1 (0.06) & $\delta\as$ \\
$R_b$ [$10^{-5}$] & 6 & $<1$ & $\delta\as$ \\
$R_\ell$ [$10^{-3}$] & 1 & 1.3 (0.7) & $\delta\as$ \\
\hline
\end{tabular}
\caption{Estimated experimental precision 
for the direct measurement of several
important electroweak precision observables at FCC-ee
\cite{fccee,cdr,d'Enterria:2016yqx} (column two, including systematic
uncertainties). Third column: parametric uncertainty of several
important EWPO due to uncertainties of input parameters given
in \eqref{eq:paraerr}, with the main source indicated in the fourth column.}
\label{tab:paraerr}
\end{table}

Note that the quoted impact of $\as$ on $R_\ell$ is a somewhat circular
statement, since $R_\ell$ is the most important pseudo-observable for the
determination of $\as$.

As discussed above, as total uncertainty for the theoretical prediction
of an observable the 
(quadratic) sum of parametric uncertainties plus intrinsic
uncertainty should be taken\footnote{It should be noted that the intrinsic
theory error is not a Gaussian random variable, which plays a role in the
combination with other error sources.}, as
given in the fourth column of Tab.~\ref{tab:futerr} and the second and
third columns of Tab.~\ref{tab:paraerr}. 
More generally, for combined fits to several observables, the parametric
uncertainties should be taken into account separately by using the
corresponding parameters in the fit. 

\bigskip
The above numbers have all been obtained assuming the SM as
calculational framework. The SM constitutes the model in which highest
theoretical precision for the predictions of EWPO can be obtained. As
soon as physics beyond the SM (BSM) will be discovered, an evaluation of the
EWPO in any preferred BSM model will be necessary.
The corresponding theory uncertainties, both intrinsic and parametric,
can then be larger
(see, e.g., \cite{futth,Heinemeyer:2004gx} for the Minimal
Supersymmetric SM).  
A dedicated theory effort (beyond the SM) would be needed in
this case.


\subsection{Higgs precision observables}

For the accurate study of the properties of the Higgs boson, precise
predictions for the various partial decay widths, the branching ratios
(BRs) and the Higgs-boson production cross sections 
along with their theoretical uncertainties are indispensable.


\subsubsection{Higgs-boson production cross-sections}
\label{hprod}

The very narrow width of the Higgs boson allows for a factorization of all
cross-sections with resonant Higgs bosons into production and decay
parts to very high precision if the Higgs boson can be fully reconstructed. 
In this case, finite-width effects and off-shell
contributions are of relative size $\Gamma_\PH/\MH \sim 0.00003$ and thus
not relevant; this is in contrast to physics with $Z$ or $W$ resonances, 
where $\Gamma/M \sim 0.03$. If the Higgs boson is not fully reconstructable
(e.g.\ in $H\to WW\to2\ell2\nu$) Higgs off-shell contributions have to be
taken into account (which is straightforward at NLO).

At FCC-ee with $\sqrt{s} = 240$~GeV (or other $e^+e^-$ machines near this
center-of-mass energy), the Higgs-boson production cross-section is strongly
dominated by $e^+e^- \to ZH$, and $e^+e^- \to \nu \bar\nu H$ contributes
less than 20\%~\cite{fccee,Moortgat-Picka:2015yla}. 
For these two processes full one-loop corrections in the SM are
available~\cite{Belanger:2002ik,Denner:2003yg}. For the dominating $ZH$
production mode they are found at the level of $\sim 5{-}10\%$. It can be
expected that missing two-loop corrections in the SM lead to an
intrinsic uncertainty of \order{1\%}%
\footnote{This estimate is corroborated
by the recent calculation of the two-loop ${\cal O}(\alpha\as)$ corrections to
$ZH$ cross-section \cite{Gong:2016jys}, which were found to amount to
1.3\%.}%
.~This number has to be compared to the anticipated experimental accuracy
of 0.4\%~\cite{fccee,cdr}. It becomes clear that with a full two-loop calculation
of $e^+e^- \to ZH$ the intrinsic uncertainty will be sufficiently
small. Calculational techniques for $2 \to 2$ processes at the two-loop
level exist, and it is reasonable to assume that, if required, this
calculation within the SM can be incorporated for the FCC-ee 
Higgs precision studies. 

For WBF production, the calculation of the full two-loop corrections will
be significantly more difficult, since this is a $2 \to 3$ process. However, one
may assume that a partial result based on diagrams with closed light-fermion
loops and top-quark loops (in a large-$\mt$ approximation) can be achieved,
which should reduce the intrinsic theory uncertainty to below the 1\% level.
Given the fact that the WBF process is less crucial than the $HZ$ channel for
the Higgs physics program FCC-ee with $\sqrt{s} = 240$~GeV,
this will probably be adequate for most practical purposes.


\subsubsection{Higgs-boson decays}
\label{hdec}

The current intrinsic uncertainties
for the various Higgs-boson decay widths are given in
Tab.~\ref{tab:Hintr}. 
They have been evaluated as follows \cite{lmp}:

\begin{table}[tb]
\centering
\renewcommand{\arraystretch}{1.1}
\begin{tabular}[t]{l|ccc}
\hline
Partial width & QCD & electroweak & total \\
\hline
$H \to b \bar b/c \bar c$ & $\sim 0.2\%$ & ${<0.3}\%$ & ${<0.4}\%$ \\
$H \to \tau^+\tau^-/\mu^+\mu^-$ & -- & ${<0.3}\%$ & ${<0.3}\%$ \\
$H \to gg$ & $\sim 3\%$ & $\sim 1\%$ & $\sim 3.2\%$ \\
$H \to \gamma\gamma$ & $< {0.1}\%$ & $< 1\%$ & ${<} 1\%$ \\
$H \to Z\gamma$ & ${\lesssim 0.1}\%$ & $\sim 5\%$ & $\sim 5\%$ \\
$H \to WW/ZZ \to 4$f & $< 0.5\%$ & ${< 0.3\%}$ & $\sim 0.5\%$ \\
\hline
\end{tabular}
\caption{
Current intrinsinc uncertainties in the various Higgs-boson decay width
calculations, see text and Refs.~\cite{lmp,lhchxswg-br}. 
}
\label{tab:Hintr}
\end{table}

The QCD uncertainty for $H\to q\bar{q}$ is assumed to be equal to the magnitude
of the ${\cal O}(\as^4)$ corrections \cite{baikov1}.  The uncertainty due to
missing ${\cal O}(\alpha^2)$ contributions is estimated to be smaller than the
known one-loop corrections \cite{hff1}, which themselves are unusually small due
to accidental cancellations. Two-loop corrections of ${\cal O}(\alpha\as)$ are
also available \cite{hbb2} and the
uncertainty from 3-loop mixed QCD-weak corrections is estimated to be of similar
size as the partial result in Ref.~\cite{hbb3}.

For $H\to gg$, the QCD uncertainty is estimated from the scale variation of the 
available N$^3$LO corrections \cite{hgg3}. The electroweak uncertainty for this
channel is estimated based on the observation that the NLO result \cite{hgg1} is
dominated by light-fermion loops, and thus the NNLO contribution is expected to
be suppressed by a factor $N_{\rm lf} \alpha \sim 0.1{-}0.2$. 
The same procedure
has been employed for $H \to \gamma\gamma$, using the results from
Ref.~\cite{haa}. Based on the experience from existing results for $H \to gg$
and $H \to \gamma\gamma$, the currently unavailable electroweak 
NLO corrections to $H\ \to Z\gamma$ are
estimated to be less than 5\%. Off-shell effects for $H\ \to Z^*\gamma$
are known at the LO one-loop level \cite{hllg} and the NLO corrections
are expected to be small compared to the experimental uncertainty.%
\footnote{
We assume that a proper experimental definition of this decay mode
w.r.t.\ Dalitz decays~\cite{hllg} will be agreed upon.
}

The
uncertainty due to the missing QCD and electroweak two-loop corrections 
for $h \to WW,ZZ$ is estimated by \emph{(i)} taking square of the
known one-loop corrections \cite{hvv1} and, alternatively, \emph{(ii)} doubling the numerical
result of the known leading two-loop corrections in the large-$\mt$ limit
\cite{hvv2}.

\begin{table}[tb]
\centering
\renewcommand{\arraystretch}{1.1}
\begin{tabular}[t]{l|ccc}
\hline
decay & para.\ $m_q$ & para.\ $\as$ & para.\ $\MH$ \\
\hline
$H \to b \bar b$ & ${1.4\%}$ & ${0.4\%}$ & {--} \\
$H \to c \bar c$ & ${4.0\%}$ & ${0.4\%}$ & {--} \\
$H \to \tau^+\tau^-$ & {--} & {--} & {--} \\
$H \to \mu^+\mu^-$ & {--} & {--} & {--} \\
$H \to gg$ & ${<0.2\%}$ & $3.7\%$ & {--} \\
$H \to \gamma\gamma$ & ${<0.2\%}$ & {--} & {--} \\
$H \to Z\gamma$ & {--} & {--} & ${2.1\%}$ \\
$H \to WW$ & {--} & {--} & ${2.6\%}$ \\
$H \to ZZ$ & {--} & {--} & ${3.0\%}$ \\
\hline
\end{tabular}
\caption{
Current parametric uncertainties in the various Higgs-boson decay width
predictions~\cite{lmp} (see text). ``--'' indicates a negligible source of
uncertainty.}
\label{tab:Hpara}
\end{table}

Also the parametric uncertainties can play a non-negligible role for the
evaluation of the partial widths. The most important parameters are the bottom quark mass
and the strong coupling constant. In Ref.~\cite{lhchxswg-br} the current
uncertainties of $\as$ and $\mb$ have been assumed to be $\delta\as = 0.0015$
and $\delta \mb = 0.03 \gev$. Additionally, we consider $\delta m_{\rm c} =
0.025\gev$, $\delta \mt = 0.85\gev$ and $\delta\MH = 0.24\gev$ \cite{pdg}. The
effect on the various partial widths has been evaluated as in Ref.~\cite{lmp} 
and is shown in Tab.~\ref{tab:Hpara}. 

When comparing the combined intrinsic and parametric uncertainties 
with the target precision of FCC-ee~\cite{fccee,cdr}%
, see Tab.~\ref{tab:fcceeh}, 
it is clear that improvements are necessary.
Concerning the intrinsic theory uncertainty, the available predictions for the
$f\bar{f}$ and $\gamma\gamma$ channels are already
sufficiently precise to match the expected FCC-ee experimental uncertainty. 
With available calculational techniques, the evaluation of complete two-loop
corrections to $H\to f\bar{f}$ can be achieved. This would reduce the
uncertainty of the electroweak contributions to less than 0.1\%. Similarly, the
complete NLO corrections to $H \to Z\gamma$ can be carried out with
existing methods, resulting in an estimated precision of about 1\%
(see above for our estimate on the Dalitz decays).

More theoretical work is needed for $H\to WW, ZZ, gg$, which are currently
limited by QCD uncertainties. 
For $H \to WW, ZZ$, the required QCD corrections are essentially
identical to those for $e^+e^- \to WW$, and as explained on page~\pageref{wwqcd}
it is straightforward to improve them to a practically negligible level.
Further
significant progress would require the calculation of two-loop electroweak
corrections, which for a $1\to 4$ process is beyond reach for the forseeable
future.

Note, however, that the $HZZ$ coupling will be mostly constrained by the
measurement of 
the $e^+e^- \to HZ$ production process at FCC-ee with
$\sqrt{s} = 240$~GeV, rather
than the decay $H \to ZZ^*$. As discussed in section~\ref{hprod}, it may be
assumed that full two-loop 
corrections (for on-shell $Z$ and $H$ bosons) will eventually be carried out for
this process, leading to a remaining intrinsic uncertainty of less than 0.3\%.

For $H \to gg$, the NNLO QCD corrections \cite{hgg2} and N$^3$LO QCD corrections in
the large-$\mt$ limit \cite{hgg3} are currently available. The leading
uncertainty stems from the missing N$^4$LO corrections in the large-$\mt$ limit.
These require the calculation of massless four-loop QCD diagrams, which may be
within reach \cite{baikov1,baikov2}. If these contributions become available,
together with three-loop corrections involving bottom loops,
the intrinsic
uncertainty for $H \to gg$ is expected to be reduced to the level of about 1\%.

\begin{table}[tb]
\centering
\renewcommand{\arraystretch}{1.1}
\begin{tabular}[t]{l|c|ccc|l}
\hline
decay & intrinsic & para.\ $m_q$ & para.\ $\as$ & para.\ $\MH$ & FCC-ee prec.\ on ${g^2_{HXX}}$ \\
\hline
$H \to b \bar b$ & $\sim 0.2\%$ & $0.6\%$ & $<0.1\%$ & -- & {$\sim 0.8\%$} \\
$H \to c \bar c$ & $\sim 0.2\%$ & $\sim 1\%$ & $<0.1\%$ & -- & {$\sim 1.4\%$} \\
$H \to \tau^+\tau^-$ & $< 0.1\%$ & -- & -- & -- & {$\sim 1.1\%$} \\
$H \to \mu^+\mu^-$ & $< 0.1\%$  & -- & -- & -- & {$\sim 12\%$} \\
$H \to gg$ & $\sim 1\%$ &  & $0.5\%$ (0.3\%) & -- & {$\sim 1.6\%$} \\
$H \to \gamma\gamma$ & $<1\%$ & -- & -- & -- & {$\sim 3.0\%$} \\
$H \to Z\gamma$ & $\sim1\%$ & -- & -- & $\sim 0.1\%$ & \\
$H \to WW$ & $\lesssim 0.3\%$ & -- & -- & $\sim 0.1\%$ & {$\sim 0.4\%$} \\
$H \to ZZ$ & $\lesssim 0.3\%^\dagger$ & -- & -- & $\sim 0.1\%$ & {$\sim 0.3\%$} \\
\hline
{$\Gamma_{\rm tot}$} & {$\sim 0.3\%$} & {$\sim 0.4\%$} 
                        & {$< 0.1\%$} & {$< 0.1\%$} & {$\sim 1\%$} \\
\hline
\multicolumn{4}{l}{$^\dagger$ From $e^+e^- \to HZ$ production}
\end{tabular}
\caption{
Projected intrinsic and parametric uncertainties for the partial and
total Higgs-boson decay width predictions (see text).
The last column shows the target of FCC-ee precisions on the respective
coupling squared. 
}
\label{tab:fcceeh}
\end{table}

Also shown in Tab.~\ref{tab:fcceeh} are the projected parametric
uncertainties, assuming FCC-ee precisions, see Tab.~\ref{tab:futerr}.
For inputs, we use $\delta\as = 0.0002$ and $\delta \mt = 50\mev$ from
eq.~\eqref{eq:paraerr}, $\delta\MH \sim 10\mev$ \cite{snowmassh}, and $\delta
\mb \sim 13\mev$ and $\delta m_{\rm c} \sim 7\mev$ \cite{lmp}. 

The corresponding uncertainties (intrinsic, parametric from quark masses,
$\as$ and $\MH$) for the total width are shown in the last line of 
Tab.~\ref{tab:fcceeh}. They are obtained by adding the uncertainties for the
partial widths linearly.  

As discussed previously, as total uncertainty of an observable the
sum of experimental, intrinsic and parametric uncertainty should be
taken.
Note that the numbers in Tab.~\ref{tab:fcceeh} do not take into account
correlations between the uncertainties in the Higgs production and decay
processes or between different decay processes, in particular entering
via $\Gamma_{\rm tot}$. Their impact can only be
evaluated when the full experimental correlation matrix is known.
Most importantly, there is a strong correlation of the parametric
uncertainty due to $\delta\mb$ between $g_{Hbb}^2$ and $\Gamma_{\rm tot}$, 
and some partial correlation of the intrinsic
uncertainties between $g_{HVV}^2$ 
($V = W, Z$) and $\Gamma_{\rm tot}$, see
Refs.~\cite{Denner:2011mq,Heinemeyer:2013tqa,lhchxswg-br} 
for more details.


\section{Conclusions}

Due to the high anticipated experimental precision at the FCC-ee  for
electroweak and Higgs-boson precision measurements, theoretical
uncertainties are expected to play an important role in the
interpretation of these measurements within the Standard Model (SM), and 
thus for expected constraints on new physics. 
We have reviewed the status of current intrinsic and parametric
uncertainties of EWPO and Higgs-boson precisison observables, which
are currently much larger than the anticipated future experimental uncertainties.
In a second step we have evaluated the possible future intrinsic and
parametric uncertainties. 

The first one can be improved through calculations of
higher-order corrections and the development of more advanced Monte-Carlo tools
in the future.
We aim to provide a guide of the possible impact and remaining uncertainty from
future multi-loop calculations. To match the anticipated FCC-ee precision, one
or two loop orders beyond what is available today will be required. It appears
promising that some progress in this direction can be made by extending existing
calculational techniques, but an extensive and concerted theory effort will be
needed to achieve the ultimate FCC-ee precision goals.
Similar considerations apply to 
other future $e^+e^-$ colliders, such as ILC, CLIC, or CEPC.

The parametric uncertainties will
improve due to the high-precision measurements of the SM input parameters
obtained at the FCC-ee itself. Other proposed $e^+e^-$ colliders have access to
subset of these measurements with larger errors; the corresponding parametric
uncertainties would have to be rescaled accordingly. 
These measurements also require that
additional radiative corrections be calculated and implemented in Monte-Carlo
tools to reduce the theory errors to the desired level.

We hope that this work helps to
facilitate the corresponding efforts.


\subsection*{Acknowledgements}

J.G.\ is supported in part by the Polish National Science Centre (NCN)
under the Grant Agreement 2017/25/B/ST2/01987 and by international
mobilities for research activities of the University of Hradec Kr\'alov\'e,
CZ.02.2.69/0.0/0.0/16\_027/0008487.
The work was partly supported by COST (European Cooperation in Science and
Technology) Action CA16201 PARTICLEFACE.
The work of S.H.\ was supported in part by the
MEINCOP (Spain) under contract FPA2016-78022-P, in part by the Spanish
Agencia Estatal de Investigaci\'on (AEI), in part by the EU Fondo
Europeo de Desarrollo Regional (FEDER) through the project
FPA2016-78645-P, in part by the ``Spanish Red Consolider MultiDark''
FPA2017-90566-REDC, and in part by the AEI through the grant IFT
Centro de Excelencia Severo Ochoa SEV-2016-0597. 
The work of S.J.\ is partly supported by
the Polish National Science Center grant 2016/23/B/ST2/03927
and the CERN FCC Design Study Programme.
T.R.\ is supported in part by an Alexander von Humboldt Honorary Research
Scholarship of the Foundation for Polish Sciences (FNP) and by the Polish
National Science Centre (NCN) under the Grant Agreement 2017/25/B/ST2/01987.
The work of C.S.\ is supported by the Heisenberg
Programme of the DFG and a fellowship of the Collaborative Research
Centre SFB 676 ``Particles, Strings, and the Early Universe'' at Hamburg
University.



\end{document}